# CURRENT AND FUTURE CAPABILITIES OF THE 74-INCH TELESCOPE OF KOTTAMIA ASTRONOMICAL OBSERVATORY IN EGYPT

Y. A. Azzam, G.B. Ali, F. Elnagahy,
H. A. Ismail, A. Haroon, I. Selim, and Ahmed-Essam

**ABSTRACT:** *In this paper, we are going to introduce the Kottamia Astronomical Observatory, KAO, to the astronomical community. The current status of the telescope together with the available instrumentations is described. An upgrade stage including a new optical system and a computer controlling of both the telescope and dome are achieved. The specifications of a set of CCD cameras for direct imaging and spectroscopy are given. A grating spectrograph is recently gifted to KAO from Okayama Astrophysical Observatory, OAO, of the National Astronomical Observatories in Japan. This spectrograph is successfully tested and installed at the F/18 Cassegrain focus of the KAO 74'' telescope.*

## INTRODUCTION

In 1905, Mr. Reynolds, an amateur astronomer at that time and later treasurer of the Royal Astronomical Society in London presented Helwan Observatory with a 30" reflecting telescope. Due to the clear sky of Helwan and the ability of the astronomers in charge of this telescope, a great deal of valuable observations was collected. As of the curiosity of astronomers to achieve more, a large telescope was necessary and a 74" telescope equipped with both Cassegrain and Coude spectrographs had been recommended to the Egyptian authorities. The Egyptian government signed a contract for this purpose with Messers. Grubb Parsons of Newcastle. U.K. in the same year (1948) that the giant 200" telescope of Mount Palomar California was erected. The delivery of the telescope and spectrographs was expected in 1955, but various difficulties arose which resulted in a considerable delay of delivery until 1963 (Samaha, 1964).

Kottamia mountain (about 80 Kms. NE of Helwan, 12 km. north of mid-point of El Maady Cairo-Ein Sukhna road, 476 meters above see-level) had been chosen for the location of Kottamia Astronomical observatory (KAO). The latitude of the observatory is 29° 55´ 35.24" N; the longitude is 31° 49´ 45.85" E. The seeing conditions prevailing at the

_________________________________________________________________
*National Research Institute of Astronomy and Geophysics, Helwan, Cairo, Egypt*

site is around 2 arc sec. in average (Hassan, 1998). KAO belongs to the National Research Institute of Astronomy and Geophysics; NRIAG (previously, Helwan Observatory).

## GENERAL OVERVIEW OF
## THE KOTTAMIA 74" TELESCOPE

The reflector is of conventional design, having a parabolic mirror with secondaries to give Newtonian, Cassegrain and Coudé foci. The mounting is of English type with north and south piers that support a polar axis with the tube offset on a short declination axis (Issa, 1986). A photo in Figure 1 shows the Kottamia telescope with its Cassegrain and Newtonian foci.

The main mirror has an aperture of 74 inches and a paraboloidal surface of 360" focal length (F/4.9). The Cassegrain secondary has an aperture of 19" and a hyperbolic section giving a focal ratio of F/18, with plate scale of 6.1 arcsec/mm.

A second hyperbolic mirror similar to the Cassegrain has a focal ratio of F/28.9 and in combination with two flats of 13.5 inches and 9.5 inches apertures give the Coude focus plate scale of 3.8 arcsec/mm.

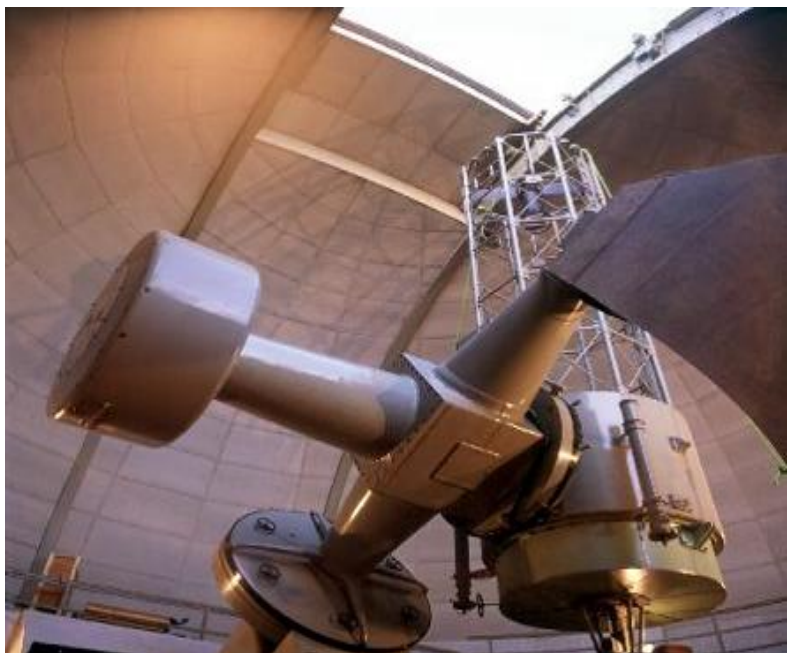

**Fig. 1: Kottamia 74" Telescope**

The Newtonian focus has a focal ratio of F/4.9, thus giving a scale of 22.53 arcsec/mm. A review of the values of focal lengths, F/ratios and plate scales at various foci of the telescope is given in Table 1.

Table 1: The Parameters of the KAO's Foci

| Focus | Focal length (m) | F-number | Plate scale (arcsec/mm) |
|---|---|---|---|
| Newtonian | 9.15 | 4.9 | 22.53 |
| Cassegrain | 33.98 | 18 | 6.09 |
| Coude | 54.29 | 28.9 | 3.8 |

## TELESCOPE DOME

The dome was constructed by the united Austrian Iron and Steel Works (VOEST) of Linz/Austria. It is 18 m in diameter with an outer shell of galvanized steel plate painted externally with aluminum. The dome is very good insulated from heat and dust which lead to diurnal range of temperature variation to be as minimum as 3 °C. A photo showing the dome is given in Figure 2.

The dome can be rotated at a speed of one revolution in six minutes by means of 6 three phase motors driving pinions into a circular rack. The dome runs on 26 traveling wheels and is kept central by means of 15 horizontal guide rollers placed around the circumference to allow for thermal expansion and distributions of wind loading. The dome has two shutters, an upper and a lower capable of moving at 1.5 m/min. Hand operation is possible in case of emergency. The upper shutter is of the "up and over" type whereas the lower shutter is in two parts each opening to the side. A canvas wind screen provide the protective action of the lower shutter. The opening of the dome is 5 meters wide and extends 2.5 meters beyond the Zenith. An observing carriage traveling at 4 m/min is fitted with control buttons for dome rotation, shutter operations and its own motion. It runs on rails attached to the main arches of the dome. Similar control pushbuttons are also provided at various positions around the periphery of the dome. At present there are no electrical control connections between the telescope and the dome.

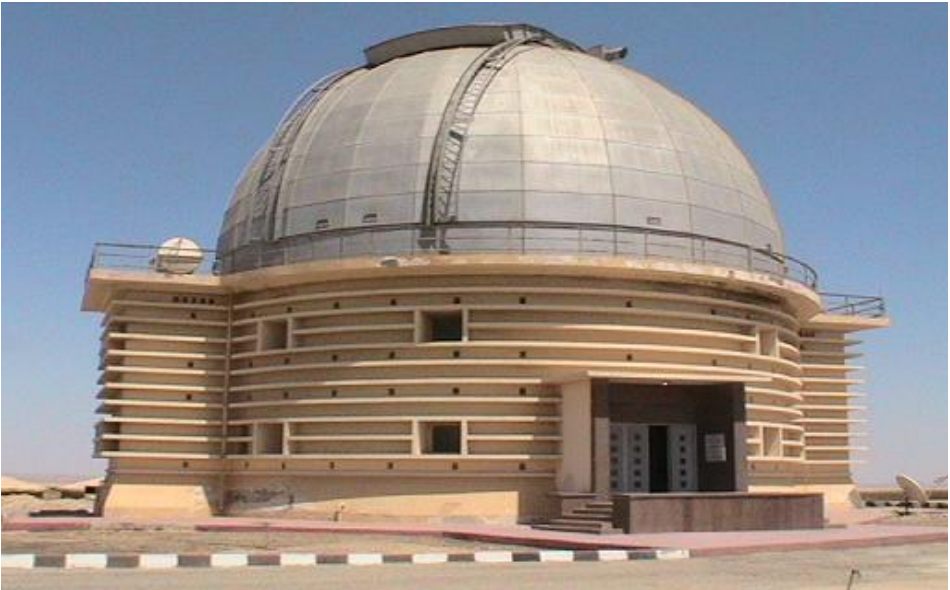

**Fig. 2: The KAO Dome**

Some few years ago, the dome control has been upgraded by one of us (F. Elnagahy) to an RF handset control. As the telescope tracks the object across the sky, it is necessary to rotate the dome from time to time to allow the light of the object to enter to the telescope. This is being done recently by the RF handset control instead of pushbuttons around the periphery of the dome, which previously necessitated a person to go up the stairs to do such movement action.

## KOTTAMIA TELESCOPE UPGRADING

In order to share astronomical communities their enormous contributions in astronomical research, it was necessary to modernize an approximately 32-years old 74" telescope. The matter was raised to governmental authorities and after some long subjective discussions, it was approved to support Kottamia telescope with the necessary money for upgrading.

**Optical System Upgrade**

In 1995, an appropriate agreement was signed with Zeiss company (FRG) to design a new optical system for the telescope. This involved a new primary mirror, secondary Cassegrain mirror both are made of

________________________________________________________________________

"Zerodur" glass ceramics, and a new supporting system for the primary mirror (Hassan, 2004). Due to some complications and problems in the design of the supporting system, the system was not approved until 2003 when it was being in effect and working properly.

The technical parameters of such optical system is as follows:

*Primary mirror:*

| Diameter | 1930 mm |
|---|---|
| Diameter of center hole | 188 mm |
| Thickness on the edge | 230 mm |
| Optical aperture | 1880 mm |
| Radius of curvature | -18277 mm |
| Material | ZERODUR |
| Thermal expansion coefficient | $0\pm0.1\times10^{-6}$/K |

*Secondary mirror*

| Diameter | 496.8 mm |
|---|---|
| Thickness in center | 80.1 mm |
| Optical aperture | 483 mm |
| Radius of curvature | 5877.62 mm |
| Conical constant | k= -3.012377 |
| Material | ZERODUR |
| Thermal expansion coefficient | $0\pm0.1\times10^{-6}$/K |

*Optical system*

| Focal ratio – Focal length | F/18 – 33984 mm |
|---|---|
| Distance M1 M2 vertex distance | 6989.9 mm |
| M1 vertex – focal plane | 1000 mm |

*Primary mirror support*

| Number of axial supports | 18 |
|---|---|
| Number of radial supports | 16 |
| Distance focal plane-instrument rotator flange | 417.5 mm |
| Maximum load onto instrument rotator flange | 400 kg with 1000 mm distance to center of gravity |

**CCD Camera System**

Before the time of the acceptance of the optical system and in 1997, the telescope has been equipped with CCD imaging system for direct imaging. The system was provided by AstroCam (later PixCellent) in U.K. This CCD system consists of two cooled CCD systems built onto an assembly that allows mounting at either the Newtonian or the Cassegrain foci of the telescope. The system has a primary CCD system consisting of a SITe (formerly Tektronics) 1024×1024 pixel (24 microns/pixel pitch) back illuminated CCD mounted in a liquid nitrogen vacuum dewar and controlled by an AstroCam 4202 controller with the Imager 2 software package. In addition, there is a second cooled CCD system that is thermoelectrically cooled. It uses a Kodak KAF-1300 CCD of 1280×1024 pixels each of 16 microns. This system is also controlled by an AstroCam 4202 CCD controller and Imager2 software. Its purpose is to act as an offset acquisition and guiding system to allow manual operation of the telescope to set it accurately to follow the science field of view without any image trailing (Osman, 2000). A photo showing the camera system is shown in Fig 3, where it is attached to the Cassegrain focus of the telescope, and detailed specifications of both cameras are given in Tables 2 and 3.

The system was enhanced to allow remote operation of all the mechanical functions that were available on the unit. This involved focusing the telescope on the sky and the guider system then brought also to be in focus. This is done with a motorized and encoded eccentric cam arrangement. The position of the acquisition field is selected remotely by the operator and a motorized slide allows the guider field of view to be moved remotely and the position of the guider displayed. The main CCD has with it a 10-position filter wheel that is encoded to show the position of the wheel on the display. A set of photometric filters (UBVRI) is inserted in the wheel (Mackay, 1995).

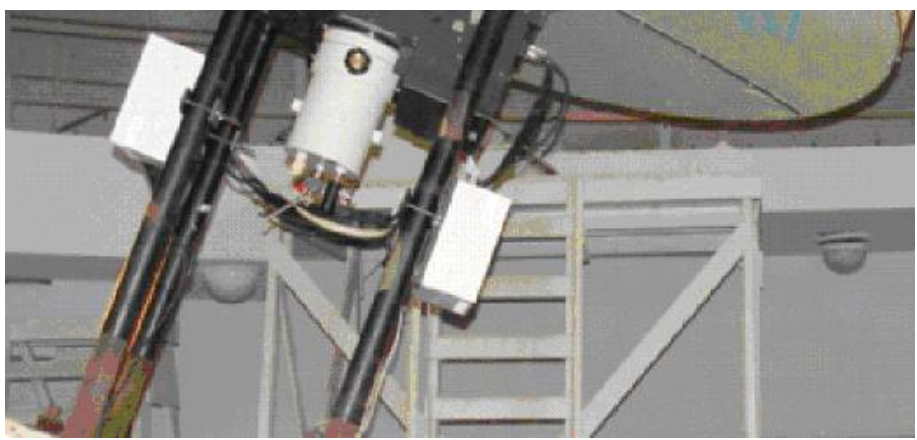

**Fig. 3: Kottamia CCD camera system attached to the Casegrain focus**

______________________________________________________________________

Table 2: Primary CCD camera system specifications

| | |
|---|---|
| Type | SITe (formerly Tektronics) |
| Format | 1024l A pixels x 1024 pixels full frame |
| Pixel size | 24 μm x 24 μm |
| Imaging area | 24.6 mm x 24.6 mm |
| Dark current @20 °C | 0.1 nA/cm$^2$ |
| Readout noise | 7 electrons |
| Full well signal | 150.000 electrons |
| Dynamic range | 20.000-30.000:1 |
| Quantum efficiency at room temperature $\lambda$ =400,500,600,700,800 nm | 60 %,70%,79%,75%,65% respectively |
| Output data rate | 50 k pixels/s |

Table 3: Secondary CCD Camera System Specifications

| | |
|---|---|
| Type | Kodak KAF-1300 L |
| Format | 1280(H) pixels x 1024(V) pixels |
| Pixel size | 16 μm x 16 μm |
| Imaging area | 20.5 mm (H) x 16.4 mm(V) |
| Dark current @25 °C | < 20 pA/cm$^2$ |
| Readout noise | 20 e$^-$ rms |
| Full well signal | 300.000 electrons |
| Dynamic range | >72 db |
| Quantum efficiency at room temperature $\lambda$ =500 nm $\lambda$ =600 nm $\lambda$ =700 nm $\lambda$ =800 nm | 25 % 35 % 42 % 48 % |
| Output data rate | 20 MHz |

## Mirror Aluminizing Plant Upgrade

In order to have high protection, efficiency and safe coating process for the new mirrors of the telescope, it was necessary to upgrade the existing aluminizing plant that was erected by EDWARDS, England in 1966. The plant was using the manual control and the obsolete analogue sensors and gauges of 1960s. As so and in 1998, the plant had an upgrading process that used BALZERS equipments. This included a new mechanical booster pump, a two stage oil sealed rotary pump, an exhaust mist filter, a higher throughput diffusion pump, a right angle baffle valve, a polycold, a backing valve, a roughing valve, a needle valve, a venting valve, a backing and roughing pipelines, and a new control cabinet. All valves used were pneumatically operated such that they all fail safe to the close positions in case of power failure, loss of water cooling supply, loss of compressed air, or pumping failure. The upgrade also included a very high efficient and intelligent cooling system, digital sensors and gauges and auto controlling all of the process by the up-to-date SIMENS Programmable Logic Control (PLC). The plant cell body, together with the filament electric power supply was the only part that did not have any change. Fig 4 shows the upgraded part of the plant.

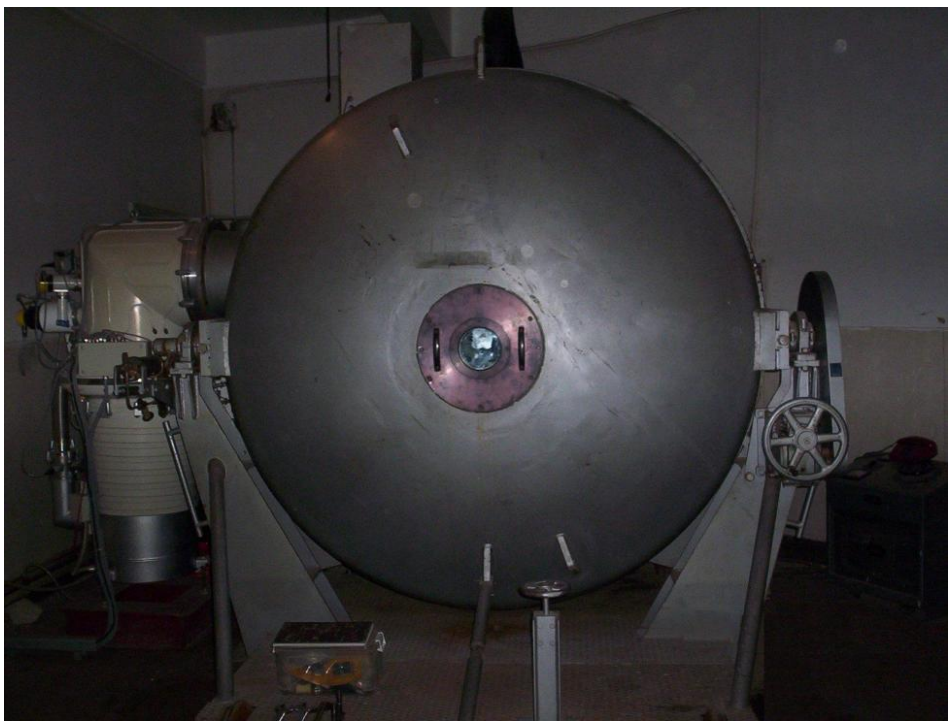

**Fig. 4: Upgraded Kottamia 74″ mirror aluminizing plant**

The new system has intelligent alarms for different expected problems related to cooling water supply cut off, reaching high temperature degrees, valves opening and closing pressures and emergency normal close of all the valves in the case of electricity failure. The system has been checked and tested thoroughly for pressure leakage, safety and for all aspects of failure operating conditions, and for the ultimate vacuum pressure that can be reached inside the plant cell. We usually start coating the mirrors by the aluminum material when the pressure reaches a value of $5\times10^{-5}$ mbar at which case the pressure inside the cell may rise slightly. The new mirror of the telescope has been successfully coated by the upgraded system and various coatings for other test mirrors have been implemented by the system.

**Okayama-Kottamia Spectrograph (OKS)**

This spectrograph was originally designed for use on the 74-inch telescope of Okayama Astrophysical Observatory (OAO) in (Japan), with an image intensifier and photographic plates. In the framework of cooperation between Japanese and Egyptian astronomers, it was suggested to present this spectrograph to be in use with the 74" telescope of KAO. The essential modification was to replace the image tube system and photographic plate by a high performance CCD camera. This modification was designed by one of us (G. B. Ali) during his visit to OAO and was implemented in OAO. The spectrograph has been brought to KAO in Sep 2006. Two Japanese staff members of OAO, came to Egypt on Feb 2007 to help KAO staff in installing and testing of the instrument. After successful completion of their mission, the spectrograph was commissioned for operation at the f/18 Cassegrain focus of the 74" telescope of KAO, see Figure 5.

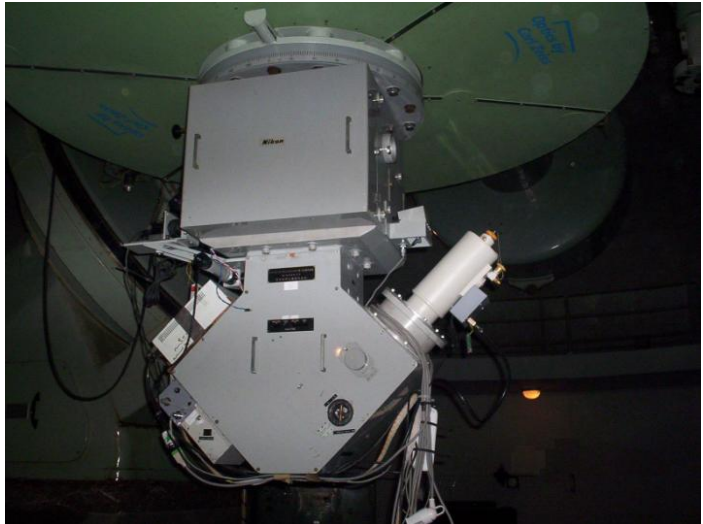

**Fig. 5: OKS Spectrograph attached to the KAO's telescope**

OKS is a low-to-moderate resolution spectrograph (700-1200). It has a slit of 15 mm length corresponding to 90 arcsec on the sky. The slit is also of variable widths ranging from 0.005 mm to 3.0 mm corresponding to 0.03 to 18 arcsec on the sky. There is also a TV system to guide off the slit or in an offset mode. The instrument has a Hollow Cathode tube (Fe + Ne) used as a comparison lamp. The main specifications of the OKS are listed in Table 4.

Table 4: Main Specifications of the OKS (Okayama-Kottamia Spectrograph)

| Collimator | | |
|---|---|---|
| | Type | inverse Cassegrain |
| | Focal length | 850mm |
| | Beam size | 47.3mm |
| | Main mirror aperture | 72 (70) mm |
| | Secondary mirror aperture | 24 (22) mm |
| | F-ratio | F/12.1 |
| Camera: | Type | Solid Schmidt |
| | Focal length | 142.2mm |
| | F-ratio | F/2.5 |
| | Field of view | 6 d |
| Slit: | Width | 5 micron ~ 3mm, one side open |
| | Length | 0.5~15mm, both sides open |
| Comparison lamp | | Hallow cathode tube (Fe+Ne) |

For the 300 gr/mm grating, the spectral resolution at different order is:

| Order | wavelength coverage over CCD resolution | |
|---|---|---|
| | **Spectral velocity** | |
| 1 | 5000A | |
| | 5.0A | 300 km/sec |
| 2 | 2500A | |
| | 2.5A | 150 km/sec |
| 3 | 1250A | |
| | 1.3A | 75 km/sec |

A rough estimation of the exposure time for a 12th magnitude star is about ten minutes at first order. Four gratings are available to provide wavelength coverage from 4000 to 8000 Å; their characteristics are given in Table 5.

Table 5: The available Gratings Characteristics

| Catalog No. | n (lines/mm) | Blaze angle | Blaze λ (Å) | Dispersion at $1^{st}$ order (Å/pixel) |
|---|---|---|---|---|
| 35-53-11-170 | 300 | 4º 18' | 5 000 | 2.84 |
| 35-63-11-640 | 300 | 10º 25' | 12 000 | 2.84 |
| 35-53-11-560 | 600 | 22º 2' | 12 500 | 1.42 |
| 35-53-11-290 | 1800 | 26º 45' | 5 000 | 0.47 |

## *Spectrograph CCD Camera System*

The CCD camera of the spectrograph is an EEV CCD 42-10 grade 1, back illuminated installed in a downward looking high performance liquid nitrogen Dewar, its specifications are given in Table 6.

________________________________________________________________

**Table 6: OKS CCD Camera Specifications**

| | |
|---|---|
| Type | EEV CCD 42-10 slow scan |
| Version | Back-illuminated (basic process), AIMO |
| Format | 2048 pixels x 512 pixels |
| Pixel size | 13.5 μm x 13.5 μm |
| Grade | 1 |
| Imaging area | 27.6 mm x 6.9 mm |
| Dark current @20 °C | 76 e/pix/s |
| @-140 °C | 0.18 e/pix/hr |
| Readout noise @83.3 KHz | 3.4 $e^-$/pixel |
| Full well signal | 96.000 electrons |
| Quantum efficiency at room temperature | |
| $\lambda$ =350 nm | 20.7 % |
| $\lambda$ =400 nm | 48.3 % |
| $\lambda$ =500 nm | 85.1 % |
| $\lambda$ =650 nm | 85.9 % |
| $\lambda$ =900 nm | 37.8 % |
| Holding time | >12 hours |

## *Test Observations*

Some of the objects are observed during and after the test phase of the OKS, Figure 6 gives some examples of the spectra for some of them as well as the spectra of the HCL comparison.

## *Suggested Scientific Programs*

Considering the advantage of KAO, i.e. the latitude, the large number of clear nights and the availability of much machine, scientific observational programs could be achieved e.g., (1) Long term monitor programs of spectral variation of variable stars, AGNs, quasars, etc. (2) Follow up observation programs of transient phenomena such as novae, supernovae, those of catastrophic binaries, etc. (3) Spectral survey programs for objects of particular kind.

a)

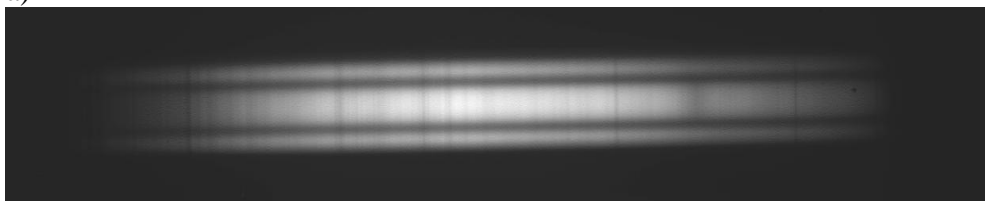

b)

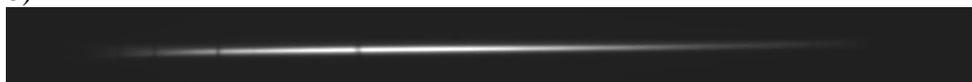

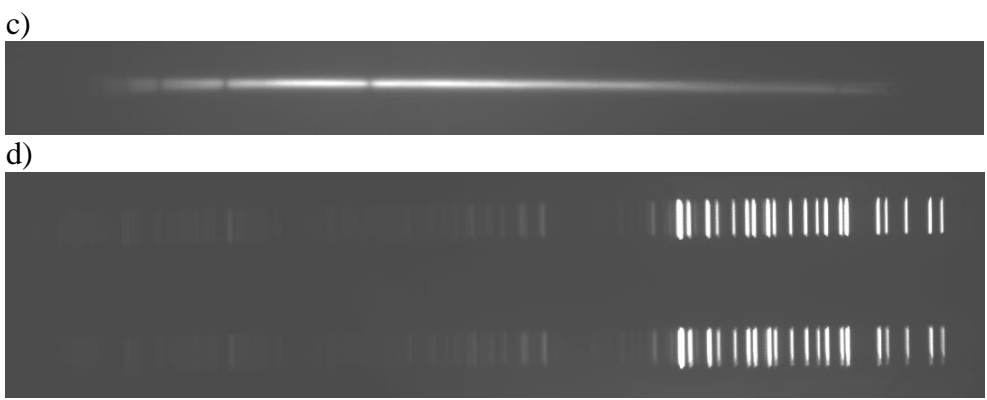

**Figure 6: Examples of test observation spectra of:**
**a) Saturn (Exp. T. is one minute)**
**b) Alpha Leo (Exp. T. is five seconds)**
**c) Alpha Crab ( Exp. T. is five seconds) and**
**d) the HCL comparison spectra of 2 second exposure tim**e.

**Telescope Control System Refurbishment**

A refurbishment process for the whole control system of the telescope is currently in effect. The refurbishment process includes replacing the existing control cabinet with the up-to-date equipment enabling computer control of all aspects of the dome and telescope. The upgrade includes also replacing of the AC motors in RA and DEC axes with high resolution stepper motors, keeping the clutch and clamp motors in both axes working for the purpose of quick/slow conversion and also for the purpose of freeing the telescope for balance. Focus motors at Cassegrain and Newtonian foci will also be replaced with high resolution stepper motors. New high resolution absolute encoders at the RA, DEC, Cassegrain and Newtonian foci will be installed. The resulting accuracy of the telescope pointing is expected to be around 5 arcsec. The accuracy in adjusting the focus position is better than ± 5 μm. An RF hand paddle will be used to move the telescope and the dome to various positions. Encoders for the dome, upper shutter, lower shutter and viewing gantry will be installed to give their positions and to allow the dome to follow the telescope in its movements. A GPS receiver will be installed for precise timing and geographical coordinates. Two rack mount industrial computers will be installed with windows XP operating system. One computer is used for the telescope control and the other one for the new purchased science CCD which will be described in the following section.

An autoguider with two stacked integrated filter wheels and a thermoelectrically cooled CCD will also be provided which has the facility to be installed in either the Cassegrain or the Newtonian foci. The software for the guider and filter wheels is fully integrated into the main telescope control system.

## Kottamia New CCD Camera System

A new CCD camera has been purchased for the purpose of direct imaging either at the Cassegrain or at the Newtonian foci. The control software of this new CCD will be integrated in the telescope control software being recently implemented. This will give an ease of operation and control of both the telescope and the camera at the same time. A new industrial computer will be dedicated to the control of the camera and saving of the scientific images. The camera is provided by Retriever Technology, USA and has the specifications detailed in Table 7.

Table 7: Specifications of the new purchased Kottamia CCD

| | |
|---|---|
| Type | EEV CCD 42-40 |
| Version | Back-illuminated with BPBC (Basic Process Broadband Coating) |
| Format | 2048 by 2048 pixels |
| Pixel size | 13.5 μm by 13.5 μm |
| Grade | 0 |
| Dynamic range | 30.000:1 |
| A/D converter | 16 bit |
| Imaging area | 27.6 mm by 27.6 mm |
| Dark current @20 °C | 0.00029 $e^-$/pix/s |
| Readout noise @20 KHz | 3.9 $e^-$/pixel |
| Full well signal | 120.000 electrons |
| Spectral Range | 200-1060 nm |
| Quantum efficiency @ -85 °C | 25.1 % @ 350 nm<br>65.5 % @ 400 nm<br>87.4 % @ 500 nm<br>80.3 % @ 650 nm<br>30.3 % @ 900 nm |
| Cooling System : | LN to -100 °C and better with temp control. Includes controller for stability within -/+ 0.1 °C. |
| Holding time | 12 hours |
| On chip binning | 1 X 1 till 8 X 8 |
| PC interface | Fiber optic |

# ACKNOWLEDGMENT

The authors would like to thank all colleagues who took part in the installation, developing and attending the tests made to various equipments of Kottamia telescope. In particular, we thank Mr El-Khamissy, Mr. Zeid, Mr. Elnaggar, and Eng. Imam. Many thanks is due to the two Japanese staff of Okayama astronomical observatory, Prof. Hiroshi Ohtani and Eng. Takafumi Okada for their great efforts and time they spent in Kottamia observatory and for their valuable explanations and discussions with the staff of Kottamia. Meanwhile we are very thankful to the director and the staff of Okayama observatory who helped in gifting and preparing the spectrograph for transfer to Kottamia Observatory.